\documentclass[letterpaper]{JHEP3}

\newcommand{\be}{\begin{equation}}
\newcommand{\ee}{\end{equation}}
\newcommand{\bear}{\begin{eqnarray}}
\newcommand{\eear}{\end{eqnarray}} 
\newcommand{\ba}{\begin{array}}
\newcommand{\ea}{\end{array}}

\newcommand{\LL}{L}
\newcommand{\vb}{\right|}

\newcommand{\CL}{{\cal L}} 
 
\newcommand{\vs}{v_{6}} 
\newcommand{\vf}{v_{4}} 
\newcommand{\etahat}{\hat{\eta}} 
\newcommand{\vev}{\hat{\Phi}_{0}} 
\newcommand{\del}{\partial}

\title{Six-Dimensional Gauge Theory on the Chiral Square 
\\ { $\; $ } \\ }

\author{Gustavo Burdman$^1$,
Bogdan A.~Dobrescu$^2$, Eduardo Pont\'{o}n$^3$ \\ \\ \\
$^1$Instituto de F\'{i}sica, Universidade de S\~{a}o Paulo, 
\\ \ \ R. do Mat\~{a}o 187, S\~{a}o Paulo, SP 05508-0900, Brazil \\ \\
$^2$Fermilab, Batavia, IL 60510, USA \\ \\
$^3$Department of Physics, Columbia University, 
 \\ \ \ 538 W. 120th St, New York, NY 10027, USA  \\ \\
\email{burdman@if.usp.br, bdob@fnal.gov, eponton@phys.columbia.edu} \\ }

\abstract{ \ \ 
We construct gauge theories in two extra dimensions compactified on
the chiral square, which is a simple compactification that leads to
chiral fermions in four dimensions.  Stationarity of the action on the
boundary specifies the boundary conditions for gauge fields.  Any
six-dimensional gauge field decomposed in Kaluza-Klein modes includes
a tower of heavy spin-1 particles whose longitudinal polarizations are
linear combinations of the extra-dimensional components, and a tower
of heavy spin-0 particles corresponding to the orthogonal
combinations.  These linear combinations depend on the Kaluza-Klein
numbers, and are independent of the gauge fixing. If 
the gauge symmetry is broken by the 
vacuum expectation value of a six-dimensional scalar, at each Kaluza-Klein 
level three spinless fields in the adjoint representation mix to provide the 
longitudinal polarization of the spin-1 mode, leaving the orthogonal 
states as two spin-0 particles.
We derive the
interactions of the Kaluza-Klein modes for generic gauge theories,
laying the groundwork for the Standard Model in two universal extra
dimensions, and more generally for future model building and
phenomenological studies.  \\ \\ }

\preprint{hep-ph/0506334 \\ 
{\small FERMILAB-Pub-05-261-T} \\ June 30, 2005 \\ } 

\keywords{\it gauge interactions; extra dimensions; compactification}

\begin{document}


\section{Introduction} \setcounter{equation}{0}

Six-dimensional (6D) gauge theories have intriguing properties that
make them potentially relevant for extensions of the standard model of
particle physics \cite{Green:1984bx}.  For example, the global
$SU(2)_W$ gauge anomaly cancels only in the case where the number of
quark and lepton generations is a multiple of three
\cite{Dobrescu:2001ae}.  Furthermore, simple compactifications of two
dimensions preserve a discrete symmetry which is a subgroup of the 6D
Lorentz group, such that the neutrino masses are forced to be of the
Dirac type and proton decay is adequately suppressed even when baryon
number is maximally violated at the TeV scale
\cite{Appelquist:2002ft}.

As with any theory that has fermions in more than four dimensions, a
major constraint imposed by the observed properties of quarks and
leptons is that the compactification allows the existence of chiral
fermions in the effective 4D theory obtained by integration over the
extra-dimensional coordinates.  The simplest compactification of
two extra dimensions, namely on a torus, does not satisfy this
constraint.  A toroidal compactification, which can be viewed as a
parallelogram with the opposite sides being identified, leads to 4D
fermions that are vector-like with respect to any gauge symmetry.  The
next simplest compactification, a parallelogram with the adjacent sides
being identified, automatically leaves at most a single 4D fermion of
definite chirality as the zero mode of any chiral 6D fermion
\cite{Dobrescu:2004zi}.  Identifying adjacent sides requires these to
have the same length $L$, so that the parallelogram has to be a rhombus
in this case.  For simplicity, we consider the most symmetric
compactification of this type: a square.  As pointed out in
\cite{Ponton:2001hq}, this configuration is naturally preferred by
radion and moduli stabilization mechanisms, a simple example of which
is the stabilization by quantum, Casimir-like effects.  We refer to
this compactification as the ``chiral square''.

In this paper we study 6D gauge theories compactified on the chiral
square.  There are various questions that we address: what are the
boundary conditions for gauge fields on the chiral square?  what kind
of gauge fixing conditions may be imposed in the 6D theory?  what is
the spectrum of Kaluza-Klein (KK) modes for gauge fields?  how does the 
Higgs mechanism work if the heavy KK gauge fields acquire masses both from
compactification and from spontaneous breaking via a vacuum
expectation value (VEV)? what are the interactions of the KK modes?
Related studies of gauge theories in six
dimensions have appeared in \cite{Csaki:2002ur, Hashimoto:2004xz,
Hashimoto:2005fe}.

Before tackling gauge fields, let us recapitulate some properties of
the chiral square derived in Ref.~\cite{Dobrescu:2004zi}.  Figure 1
shows a chiral square, with adjacent sides identified.  This
space has the topology of a sphere, but has a flat metric except
for conical singularities at $(0,0)$ and $(L,L)$ (deficit angle of
$3\pi/2$ each) and a third one at the identified points $(0,L) \sim
(L,0)$ (deficit angle of $\pi$).
\begin{figure}
\begin{center}
\begin{picture}(-340,150)(200,-50)
\thinlines
\put(-50,-40){\vector(1, 0){156}}
\put(-40,-50){\vector(0, 1){140}}
\thicklines
\put(-40,-41){\line(1, 0){80}}
\put(-40,-40){\line(1, 0){80}}
\put(-40,-39){\line(1, 0){80}}
\put(-41,-40){\line(0, 1){80}}
\put(-40,-40){\line(0, 1){80}}
\put(-39,-40){\line(0, 1){80}}
\put(40,40){\line(-1, 0){80}}
\put(40,38){\line(-1, 0){80}}
\put(38,40){\line(0, -1){80}}
\put(40,40){\line(0, -1){80}}
\put(-50,-50){$0$}
\put(-54,38){$L$}
\put(38,-52){$L$}
\put(100,-52){$x^{4}$}
\put(-54,87){$x^{5}$}
%
\end{picture}
\end{center}
\caption{The chiral square: the sides marked by thick lines are
identified, and the sides marked by a double line are also identified.
}
\label{schannelfig}
\end{figure}
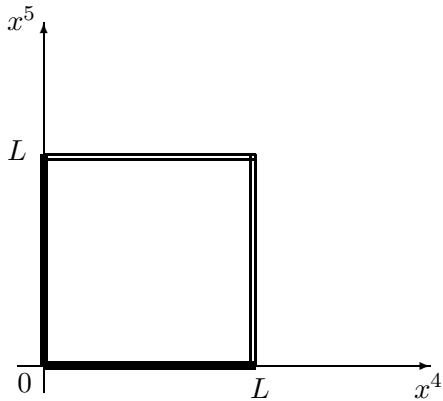
From this point of view there is nothing special going on at the sides
that are being glued together.  It is nevertheless useful to formulate
this compactification by considering the above square region on the
$x^4-x^5$ plane, together with boundary conditions on the edges of the
square that encode the identification of adjacent sides.  In particular, field values
should be equal at identified points, modulo possible symmetries of
the theory.  The physics at identified points is identical if the
Lagrangians take the same values for any field configuration:
\bear
\CL(x^\mu, y,0) = \CL(x^\mu, 0, y) ~,
\nonumber \\ [.3em] 
\CL(x^\mu, y,\LL) = \CL(x^\mu, \LL, y) ~,
\eear
for any $y \in [0,L]$.  For a free field $\Phi$, this requirement is
consistent with
\be
\Phi(x^\mu, y, 0)  =  e^{i \theta} \Phi(x^\mu, 0, y) ~,
\label{scalar-bc}
\ee
for an arbitrary phase $\theta$, provided one requires a ``smoothness'' 
condition on the derivatives normal to the ``edges'' of the square
\bear
\left.\partial_5\Phi \vb_{(x^4, x^5)=(y,0)}  = 
- e^{i  \theta} 
\left.\partial_4\Phi \vb_{(x^4, x^5)=(0,y)} ~.
\label{scalar-bc2}
\eear
Similar relations should be imposed at $(L,y) \sim (y,L)$.  However,
it was found in \cite{Dobrescu:2004zi} that this system admits
nontrivial solutions only when $\theta$ takes one of the four discrete
values $n \pi/2$, for $n=0,1,2,3$.  Only those fields that satisfy
boundary conditions corresponding to $n=0$ admit a zero-mode.
Furthermore, when considering 6D Weyl fermions, one finds that their
4D left- and right-handed chiralities obey boundary conditions
corresponding to integers that differ by one: $n_{L} - n_{R} = \pm 1$,
where the sign depends on the 6D chirality.  Hence, fermions
propagating on this space lead necessarily to a chiral low-energy
theory: at most one of the left- or right-handed chiralities has a
zero mode. This compactification is equivalent to the $T^{2}/Z_{4}$
orbifold \cite{Dobrescu:2004zi}.

The chiral square possesses a discrete $Z_{8}$ symmetry that acts on
the right- and left-handed components of 6D Weyl spinors as
\bear
\label{componentZ8symmetry}
\Psi_{\pm R}(x^\mu, x^4, x^5) &\mapsto& e^{-i (n^{\pm}_{R} \pm 1/2)
\pi/2} \Psi_{\pm R}(x^\mu, x^4, x^5) ~,
\nonumber \\
\Psi_{\pm L}(x^\mu, x^4, x^5) &\mapsto& e^{-i (n^{\pm}_{L} \mp 1/2)
\pi/2} \Psi_{\pm L}(x^\mu, x^4, x^5) ~,
\eear
where $+$ or $-$ label the 6D chirality, and $n^{\pm}_{L}$, $n^{\pm}_{R}$ 
label the boundary conditions satisfied by
$\Psi_{\pm L}$ and $\Psi_{\pm R}$, respectively.  Note that each KK tower
transforms as a single entity under the $Z_{8}$, i.e. the symmetry commutes
with KK number.  This $Z_{8}$ symmetry is at the heart of the Dirac
nature of neutrinos and the suppression of baryon number violation.

A KK-parity can also be naturally imposed on these scenarios and
ensures that the lightest KK particle (LKP) is a viable dark matter
candidate \cite{Servant:2002aq}.
This $Z_{2}$ symmetry distinguishes among KK modes and acts as 
\be
\label{KKparity}
\Phi^{(j,k)} (x^\mu) \mapsto (-1)^{j+k} \, \Phi^{(j,k)} (x^\mu) ~,
\ee
where $\Phi$ stands for a field of any spin, and $j,k$ are integers
labeling the KK level.  The KK-parity has a geometrical interpretation
as a rotation by $\pi$ about the center of the chiral square.  In
particular, KK-parity requires that localized operators at $(0,0)$ and
$(L,L)$ be identical.  Localized operators at $(0,L) \sim (L,0)$
have coefficients that are, in general, unrelated to those on the
previous two conical singularities.

In section 2 we give the appropriate boundary conditions for gauge
fields propagating on the chiral square.  We concentrate on those
boundary conditions that preserve a zero mode, i.e. we do not study
the breaking of gauge symmetries by boundary conditions.  Next we turn
to deriving the self-interactions of the KK modes in the mass
eigenstate basis for non-Abelian gauge fields (section 3), gauge
interactions of fermions (section 4) and scalars (section 5).  In
section 6 we analyze spontaneously broken 6D gauge symmetries.  We
also address in detail the gauge fixing procedure, with and without
breaking by nonzero VEV's, discuss the associated ghost action and
corresponding KK decomposition, isolate the linear combinations of
scalars that provide the longitudinal polarizations for massive gauge
fields, and identify the additional scalars coming from the extra
dimensional components of the gauge fields.  We summarize and conclude
in section 7.

\bigskip

\section{Abelian gauge fields}
\label{gaugebcs}
\setcounter{equation}{0}

Let us first study a 6D Abelian gauge field, $A^\alpha (x^\beta)$ with
$\alpha, \beta = 0,1,...5$, whose propagation in the $x^4,x^5$ plane
is restricted to a square of size $L$ ($x^\nu$, $\nu = 0,1,2,3$ are
the Minkowski spacetime coordinates).  The action has the usual
quadratic form in the gauge field strength, $F^{\alpha\beta}$,
\be
S = \int\,d^4x \int_{0}^{L}dx^4 \int_{0}^{L} dx^5\,
\left(-\frac{1}{4}F_{\alpha\beta}F^{\alpha\beta} + {\cal L}_{GF}\right)~,
\label{s1}
\ee
and includes a gauge fixing term, ${\cal L}_{GF}$, which we choose
such that the mixings of $A_\mu$, $\mu = 0,1,2,3$, with $A_4$ and
$A_5$ vanish:
\be
{\cal L}_{GF}=-\frac{1}{2\xi}
\left[\frac{}{} \partial_\mu A^\mu - \xi\left(\partial_4A_4 + 
\partial_5A_5 \right)\right]^2~,
\label{gf}
\ee
where $\xi$ is the gauge fixing parameter. 

The action may also include localized kinetic terms at the fixed
points $(0,0)$, $(L,L)$ and $(0,L)\sim (L,0)$.  Such terms are
generated radiatively, as was explicitly shown in the cases of 5D
theories \cite{Georgi:2000ks} and of 6D theories compactified on the
$T^2/Z_2$ orbifold \cite{Cheng:2002iz}, and are phenomenologically
important \cite{Cheng:2002ab,Carena:2002me}.  In this work we assume
that they are sufficiently small that they can be taken into account
perturbatively.  Therefore, Eq.~(\ref{s1}) is our starting point for
the KK decomposition.  We defer a detailed study of localized terms
for future work \cite{us1}.

\medskip

\subsection{Boundary conditions}

Integrating by parts, the action (\ref{s1}) can be written as
\bear
S & = & 
\int\,d^4x \int_{0}^{L}dx^4 \int_{0}^{L} dx^5\,\left\{
-\frac{1}{4}F_{\mu\nu}F^{\mu\nu} -\frac{1}{2\xi}(\partial_\mu A^\mu)^2
+\frac{1}{2}\left[ (\partial_4 A_\mu)^2  + (\partial_5 A_\mu)^2 \right]
\right.\nonumber\\ [0.5em]
&&  \left. \mbox{} + \frac{1}{2} \left[ (\partial_\mu A_4)^2 
+ (\partial_\mu A_5)^2 - \xi (\partial_4 A_4 + \partial_5 A_5)^2 -
(\partial_4 A_5 - \partial_5 A_4)^2 \right] 
\right.\nonumber\\ [0.5em]
&&  \left. \mbox{} + \partial_4 \left[ A_4 \partial_\mu A^\mu \right]
+ \partial_5 \left[ A_5 \partial_\mu A^\mu \right] \rule{0mm}{5.8mm} \right\}~.
\label{s2}
\eear
The last two terms are surface terms that are important in determining
the possible boundary conditions.

The variation of $S$ with respect to $A^\alpha$ must vanish everywhere
in the bulk, leading to the following field equations:
\bear 
\partial^\mu F_{\mu\nu}+\frac{1}{\xi}\partial_\mu\partial_\nu A^\mu
& = & \left( \partial_4^2 + \partial_5^2 \right) A_\nu ~,
\nonumber 
\\ [0.5em]
\left(\partial_\mu\partial^\mu - \xi \partial_4^2
-\partial_5^2 \right) A_4 & = & (\xi - 1)\partial_4\partial_5 A_5 ~,
\label{a4eom}
\\ [0.6em]
\left(\partial_\mu\partial^\mu -\partial_4^2
- \xi \partial_5^2 \right) A_5 & = & (\xi - 1)\partial_4\partial_5 A_4 ~.
\nonumber 
\eear
Furthermore, we require the surface terms in $\delta S$ to vanish
everywhere on the boundary:
\bear 
&& \hspace*{-5em} \int\,d^4x \left\{ \int_{0}^{L}dx^4 
\left.\left[ \frac{}{} F_{5\mu} \delta A^\mu
+ F_{45} \delta A_4 + \left( \partial_\mu A^\mu - \xi \partial_4 A_4
- \xi \partial_5 A_5 \right) \delta A_5 \right]\right|_{x^5=0}^{x^5=L} 
\right.
\nonumber \\ [0.7em]
&& \hspace*{-3em} \left. + \int_{0}^{L}dx^5 \left. \left[ \frac{}{} 
F_{4\mu} \delta A^\mu
- F_{45} \delta A_5 + \left( \partial_\mu A^\mu - \xi \partial_4 A_4
- \xi \partial_5 A_5 \right) \delta A_4 \right]\right|_{x^4=0}^{x^4=L} 
\; \right\}
= 0 ~.
\label{bc1}
\eear
This leads to boundary conditions that guarantee a well-defined,
self-adjoint problem, which in turn ensures that the differential
operators in Eqs.~(\ref{a4eom}) possess a complete set of orthogonal
eigenfunctions.  Possible localized terms in the original action would
give additional contributions to Eq.~(\ref{bc1}), and thus would
correspond to a modification of the boundary conditions.  Requiring
that the boundary contributions to $\delta S$ vanish also guarantees
that there is no flow of charges, such as energy or momentum, across
the boundary, or that any flow is explicitly associated with localized
terms that act as sources for the corresponding charges.  As already
mentioned, we do not consider localized terms in what follows.

As discussed in Section 1, we consider the case where the points
$(y,0)$ and $(0,y)$ are identified in the sense that the Lagrangians
at these points are equal, and likewise $(y,L)$ and $(L,y)$ are
identified, for any $0\le y \le L$.  Given that any matter field
$\Phi$ satisfies the boundary conditions (\ref{scalar-bc}) and
(\ref{scalar-bc2}), and analogous relations at the boundaries
$x^{4}=L$ and $x^{5}=L$, the requirement that the boundary conditions
are compatible with the gauge symmetry implies
\bear
&& \left.D_\mu\Phi \vb_{(x^4, x^5)=(y,0)}  = 
e^{i \theta} \left.D_\mu \Phi \vb_{(x^4, x^5)=(0,y)}~, 
\nonumber  \\ [.3em]
&& \left.D_4\Phi \vb_{(x^4, x^5)=(y,0)}  = 
e^{i  \theta} 
\left.D_5 \Phi \vb_{(x^4, x^5)=(0,y)}~, 
\nonumber  \\ [.3em]
&& \left. D_5\Phi \vb_{(x^4, x^5)=(y,0)}  = 
- e^{i  \theta} 
\left. D_4\Phi \vb_{(x^4, x^5)=(0,y)} ~.
\label{covariant-bc}
\eear
The first and second equations are derived by differentiating
Eq.~(\ref{scalar-bc}) with respect to $x^\mu$ and $y$, respectively,
and then replacing partial derivatives by covariant ones,
\be
\label{covariant}
D_{\alpha} = \partial_{\alpha} - i g_6 A_{\alpha} ~,
\ee 
where the 6D gauge coupling, $g_6$, has mass dimension $-1$.  The last
equation in (\ref{covariant-bc}) is obtained directly from
Eq.~(\ref{scalar-bc2}).

Eq.~(\ref{covariant-bc}) implies that the boundary conditions are
invariant under 6D gauge transformations only if $A_\mu$, $A_4$ and
$A_5$ satisfy the ``folding'' identifications
\bear
A_\mu(y,0)&=&A_\mu(0,y)~,\nonumber\\
A_4(y,0)&=&A_5(0,y)~,\label{foldingBC}\\
A_5(y,0)&=&-A_4(0,y)~,\nonumber
\eear
and the same relations between fields at $(y,L)$ and $(L,y)$. These boundary conditions have also been derived in \cite{Hashimoto:2005fe}.

Alternatively, one can understand the sign in the last equation of
(\ref{foldingBC}) by recalling that the folding boundary condition
(\ref{scalar-bc}) is closely related to rotations by $\pi/2$ about the
origin of the larger square $-L < x^{4}, x^{5} < L$.  Under $(x^{4},
x^{5}) \mapsto (-x^{5}, x^{4})$, the gauge field satisfies the
covariant transformation law $(A_{4}, A_{5}) \mapsto (A_{5}, -A_{4})$,
and identifying boundary points that differ by such a rotation leads
to Eq.~(\ref{foldingBC}).

In the presence of the folding identifications (\ref{foldingBC}),
Eq.~(\ref{bc1}) implies that either the gauge field values are fixed
on the boundary ($\delta A_\alpha = 0$), or else
\bear
F_{\mu5}(y,0) &=& - F_{\mu4}(0,y)~,\nonumber \\[0.5em]
F_{45}(y,0) &=& F_{45}(0,y)~,\label{bc2} \\[0.6em]
\left[\frac{}{} \partial_\mu A^\mu - \xi \left(\partial_4 A_4 + 
\partial_5 A_5\right)
\right]_{(x^4,x^5)=(y,0)} &=&
\left[\frac{}{} \partial_\mu A^\mu - \xi 
\left(\partial_4 A_4 + \partial_5 A_5\right)\right]_{(x^4,x^5)=(0,y)}~.
\nonumber
\eear
%
In the latter case, differentiating Eqs.~(\ref{foldingBC}) with
respect to $y$ and combining with Eq.~(\ref{bc2}) we find some
constraints on the $x^4$ and $x^5$ derivatives of $A_\alpha$ on
adjacent sides of the square.  For $A_\mu$, the full set of boundary
conditions reads
\bear
A_\mu(y,0) &=& A_\mu(0,y)~,\nonumber \\ [0.4em]
\partial_4 A_\mu|_{(x^4,x^5)=(y,0)} &=& \partial_5 
A_\mu|_{(x^4,x^5)=(0,y)}~,\label{Amubc} \\ [0.5em]
\partial_5 A_\mu|_{(x^4,x^5)=(y,0)} &=& -\partial_4 
A_\mu|_{(x^4,x^5)=(0,y)}~,\nonumber
\eear
and analogous relations at $(y,L)$ and $(L,y)$.  The conditions on the
$A_{4}$ and $A_{5}$ components of the gauge field are more
conveniently expressed in terms of the fields $A_\pm = A_4 \pm i A_5$:
\bear
A_\pm(y,0) &=& \mp \, i A_\pm(0,y)~,\nonumber \\ [0.5em]
\partial_4 A_\pm|_{(x^4,x^5)=(y,0)} &=& \mp \, i \,\partial_5 
A_\pm|_{(x^4,x^5)=(0,y)}~,\label{Bpmbc} \\ [0.5em]
\partial_5 A_\pm|_{(x^4,x^5)=(y,0)} &=& \pm \, i\, \partial_4 
A_\pm|_{(x^4,x^5)=(0,y)}~.\nonumber
\eear

\medskip

\subsection{Kaluza-Klein decomposition}

Without loss of generality, we may expand the fields $A_{\mu}$,
$A_{\pm}$ in terms of complete sets of functions satisfying the
boundary conditions (\ref{Amubc}) and (\ref{Bpmbc}).  Using the
complete sets of functions satisfying these boundary conditions found
in \cite{Dobrescu:2004zi}, we may write
\bear
A_\mu(x^\nu,x^4,x^5)&=& \frac{1}{L}\left[ A_\mu^{(0,0)}(x^\nu) 
+ \sum_{j\ge 1} \sum_{k\ge 0} 
f_0^{(j,k)}(x^4, x^5) A_\mu^{(j,k)}(x^\nu) \right]
~, \nonumber\\ [0.4em]
A_+(x^\nu,x^4,x^5) &=& 
-\frac{1}{L}\sum_{j\ge 1} \sum_{k\ge 0} f_3^{(j,k)}(x^4, x^5) 
A_+^{(j,k)}(x^\nu) ~,
\nonumber\\ [0.4em]
\label{KKgaugepm}
A_-(x^\nu,x^4,x^5) &=& 
\frac{1}{L}\sum_{j\ge 1} \sum_{k\ge 0} f_1^{(j,k)}(x^4, x^5) 
A_-^{(j,k)}(x^\nu) ~,
\eear
where $j$ and $k$ are integers, $A_\mu^{(j,k)}$ and $A_\pm^{(j,k)}$
are canonically normalized KK modes, and the KK functions $f_n$ with
$n=0,1,2,3$ are: 
\bear
\label{KKsolns}
&& f_{0,2}^{(j,k)}(x^4, x^5) = \frac{1}{1+\delta_{j,0}}
\left[\cos \left( \frac{j x^4 + k x^5}{R} \right)
\pm \cos \left( \frac{k x^4 - j x^5}{R} \right)\right] ~, 
\nonumber \\ [.8em] 
&& f_{1,3}^{(j,k)}(x^4, x^5) =  i \sin\left( \frac{j x^4 + k x^5}{R} \right)
 \mp \sin\left( \frac{k x^4 - j x^5}{R} \right)  ~.
\eear
These satisfy the two-dimensional Klein-Gordon equation,
\bear
\left( \partial_{4}^{2} + \partial_{5}^{2} + M^2_{j,k} \right) 
f_{n}^{(j,k)}(x^4, x^5) = 0 ~,
\eear
where
\be
M^2_{j,k}\equiv \frac{j^2+k^2}{R^2}~,
\label{massdef}
\ee
with $R= L/\pi$, and are normalized so that
\be
\label{ortho}
\frac{1}{L^2}
\int_{0}^{\LL} dx^4 \int_{0}^{\LL} dx^5
\left[ f_{n}^{(j,k)}(x^4, x^5) \right]^* f_{n}^{(j^\prime,k^\prime)}(x^4, x^5)
= \delta_{j,j^\prime}\, \delta_{k,k^\prime}~.
\ee
Note that $f_{1}^{(j,k)} = - f_{3}^{(j,k)*}$, so that the explicit
minus sign in the expansion of $A_+$ shown in Eq.~(\ref{KKgaugepm})
leads to $A_{-}^{(j,k)} = A_{+}^{(j,k)*}$.  Derivatives along $x^4$ or
$x^5$ acting on the KK functions satisfy
\be
\partial_{\pm} f_{n}^{(j,k)}(x^4, x^5)
= i r_{j,\pm k} M_{j,k} f_{n \mp 1}^{(j,k)}(x^4, x^5) ~,
\label{delf}
\ee
where 
\be
\partial_{\pm} = \partial_{4} \pm i\partial_{5}
\label{partialpm}
\ee
and $r_{j,k}$ are 
complex phases that depend only on the KK-numbers:
\be
r_{j,k} \equiv \frac{j + i k}{\sqrt{j^2+k^2}} ~.
\label{rjk}
\ee

Before inserting the KK expansions (\ref{KKgaugepm}) into the action
(\ref{s2}), note that 
\bear 
\partial_{4} A_{5} - \partial_{5} A_{4} &=& \frac{1}{L}\sum_{j\ge 1} 
\sum_{k\ge 0} M_{j,k} \, A_H^{(j,k)}(x^\nu) f_{0}^{(j,k)}(x^4, x^5)~,
\nonumber \\
\partial_{4} A_{4} + \partial_{5} A_{5} &=& 
\frac{1}{L}\sum_{j\ge 1} \sum_{k\ge 0} M_{j,k} \, A_G^{(j,k)}(x^\nu) 
f_{0}^{(j,k)}(x^4, x^5)~,
\label{goldstone}
\eear 
where we defined at each KK level two real scalar fields,
$A_H^{(j,k)}$ and $A_G^{(j,k)}$, by
\be
A_{\pm}^{(j,k)} = r_{j,\pm k} \left(A_H^{(j,k)} \mp i A_G^{(j,k)}\right)~.
\label{hphi}
\ee
The explicit factors of $M_{j,k}$ in Eqs.~(\ref{goldstone}) ensure
that the scalars $A_H^{(j,k)}$ and $A_G^{(j,k)}$ are canonically
normalized.  It is clear that $A_H^{(j,k)}$ correspond to excitations
which are invariant under 6D gauge transformations: $A_{\alpha}
\rightarrow A_{\alpha} + \partial_{\alpha} \chi/g_{6}$, where $\chi$
is a gauge parameter that, like $A_{\mu}$ in Eq.~(\ref{KKgaugepm}),
has an expansion in terms of $f_{0}^{(j,k)}$.  The orthogonal
excitations, $A_G^{(j,k)}$, shift under such a gauge transformation
and correspond to the Nambu-Goldstone modes eaten by the massive 4D
fields, $A_\mu^{(j,k)}$.

After integrating over $x^4$ and $x^5$, we find the 4D Lagrangian for
the KK modes (one can check that the last two terms in Eq.~(\ref{s2})
give no contribution).  The gauge boson of level $(0,0)$ remains
massless, while the gauge bosons of level $(j,k)$ with $j\ge 1$ appear
as massive vector particles in an $R_\xi$ gauge, with a Lagrangian
\be
 -\frac{1}{4}F^{(j,k)}_{\mu\nu} F^{(j,k)\mu\nu}
+\frac{1}{2} M^2_{j,k}\left(A_\mu^{(j,k)}\right)^2 
-\frac{1}{2\xi}\left(\del^\mu A^{(j,k)}_\mu\right)^2 ~,
\label{spin_one_action}
\ee
where $F_{\mu\nu}^{(j,k)}=\del_\mu A^{(j,k)}_\nu - \del_\nu
A^{(j,k)}_\mu$ is the 4D field strength of level $(j,k)$, and for the
zero-mode one just sets $M_{0,0} = 0$.  At each $(j,k)$ level with
$j\ge 1$ one finds that $A_H^{(j,k)}$ and $A_G^{(j,k)}$, as defined in
Eqs.~(\ref{goldstone}) and (\ref{hphi}), are mass eigenstates in the
gauge defined by Eq.~(\ref{gf}), and are described by the following
terms in the 4D Lagrangian:
\be
\frac{1}{2}\left[ \left(\del_\mu A_H^{(j,k)}\right)^2 -M^2_{j,k} 
\left(A_H^{(j,k)}\right)^2
+\left(\del_\mu A_G^{(j,k)}\right)^2 -\xi
M^2_{j,k} \left(A_G^{(j,k)}\right)^2\right] ~.
\label{l4dfin}
\ee
One can explicitly check that the field equations (\ref{a4eom}), when
expressed in terms of $A_H^{(j,k)}$ and $A_G^{(j,k)}$, are satisfied.
In the unitary gauge, $\xi \rightarrow \infty$, only the
$A_H^{(j,k)}(x)$ scalars propagate with masses $M_{j,k}$, whereas the
fields $A_G^{(j,k)}(x)$ are the Nambu-Goldstone bosons eaten by the
$A^{(j,k)}_\mu(x)$ KK gauge bosons.

\bigskip

\section{Non-Abelian gauge fields}
\setcounter{equation}{0}

The boundary conditions, KK decomposition, and identification of the
physical states in the case of non-Abelian gauge fields are analogous
to the ones discussed in Section 2 for Abelian fields.  In this
section we present the self-interactions of the KK modes associated
with non-Abelian gauge fields, and then we study the ghost fields
required by gauge fixing.

\medskip

\subsection{Self-Interactions}

The kinetic term of a 6D non-Abelian gauge field, $A^a_\alpha$, where
$a$ labels the generators of the adjoint representation, is given by
\be
-\frac{1}{4} F^a_{\alpha\beta}F^{a\, \alpha\beta} = 
-\frac{1}{4} \left( F^{a}_{\mu\nu} F^{a\, \mu\nu} - 
2 F^{a}_{+\mu} F_{-}^{a\mu} \right)
+ \frac{1}{8} \left( {F^{a}_{+-}} \right)^{2} ~.
\label{non-abelian}
\ee
The gauge field strengths introduced here are defined by 
\bear
&& F^a_{\mu\nu} = \partial_{\mu} A_{\nu}^a - \partial_{\nu} A^a_{\mu} 
+ g_6 f^{abc} A^b_{\mu} A^c_{\nu}~,
\nonumber \\ [0.3em]
&& F^a_{\pm\mu} = \partial_{\pm} A^a_{\mu} - \partial_{\mu} A^a_{\pm} 
+ g_6 f^{abc} A^b_{\pm} A^c_{\mu} = F^a_{4\mu} \pm i F^a_{5\mu}~,
\nonumber \\ [0.3em]
&& F^a_{+-} = \partial_{+} A^a_{-} - \partial_{-} A^a_{+} 
+ g_6 f^{abc} A^b_{+}A^c_{-} = -2 i F^a_{45} ~,
\eear
where $f^{abc}$ are the group structure constants.

The trilinear terms included in Eq.~(\ref{non-abelian}) are given by
\be
-\frac{g_6}{4} f^{abc} \left\{ 4 A^{a\,\mu} A^b_\nu \, \partial_{\mu} A^{c\,\nu} 
+ \left[ A_-^a A_+^b \,\partial_+ A_-^c 
+ 2  A_-^a A^{b\,\mu} \left( \partial_{\mu} A_+^c - \partial_{+} A^c_\mu\right)
 + {\rm H.c.} \right] \right\} ~.
\ee
Upon KK decomposition and integration over the $x^4$ and $x^5$
coordinates, these give rise in the 4D Lagrangian to trilinear
interactions (proportional to the 4D gauge coupling, $g_{4} =
g_{6}/L$) among spin-1 modes,
\be
\label{tri-gauge}
- g_4 f^{abc} \delta_{0,0,0}^{(j_{1}k_{1})(j_{2}k_{2})(j_{3}k_{3})} 
 A^{(j_{1},k_{1})\,a}_\mu  A^{(j_{2},k_{2})\,b}_\nu
\,\partial^{\mu}\!A^{(j_{3},k_{3})\,c\,\nu}~,
\ee
as well as trilinear interactions involving one, two or three scalar
gauge fields:
\bear
\label{tri-scalar}
\frac{g_4}{2} f^{abc} 
\delta_{1,3,0}^{(j_{1}k_{1})(j_{2}k_{2})(j_{3}k_{3})} 
A^{(j_1,k_1)\, a}_- && \hspace{-.5em} \left[ 
-i r_{j_2,k_2} M_{j_2,k_2} A^{(j_2,k_2)\,b}_{\mu}  A^{(j_3,k_3)\,c\,\mu} 
- \left(\partial^\mu A^{(j_2,k_2)\,b}_+\right) A^{(j_{3},k_{3})\, c}_\mu 
 \right.
\nonumber \\ [0.3em]
&& \left. \mbox{} + \frac{i}{2} r_{j_3,k_3} M_{j_3,k_3} 
A^{(j_2,k_2) \, b}_+ A^{(j_3,k_3)\,c}_-
\right] + {\rm H.c.} 
\eear
Here we used Eq.~(\ref{delf}) to express the $\partial_{\pm}$
derivatives in terms of the KK masses.

We have introduced the following notation:
\be
\label{selection}
\delta_{n_{1},...,n_r}^{(j_1, k_1) ... (j_r, k_r)} \equiv 
\frac{1}{L^2}\int_{0}^{\LL} dx^4 \int_{0}^{\LL} dx^5 \,
f_{n_1}^{(j_1,k_1)} ... f_{n_r}^{(j_r,k_r)} ~.
\label{deltadefn}
\ee
This integral was computed for $r=3$ in Ref.~\cite{Dobrescu:2004zi},
and the result in the particular cases relevant for trilinear
interactions read
\bear 
&& \hspace*{-.5em}
\delta_{n,\overline{n},0}^{(j_1, k_1)(j_2, k_2 )( j_3, k_3 )}
 = \frac{ 1 }{2 \left(1+\delta_{j_1, 0} \right) \left(1+\delta_{j_2, 0} \right)
  	\left(1+\delta_{j_3, 0} \right) }
\; \left[ \rule{0em}{1em}
7 \delta_{j_1, 0}\delta_{j_2, 0} \delta_{j_3, 0}
+ \delta_{j_1, j_3 - j_2}\delta_{k_1, k_3-k_2} 
\right.
\nonumber \\ [0.7em]
&& \hspace*{0.8em} + \, i^n \left(  \delta_{j_1, j_3 + k_2}
        \delta_{k_1, k_3 - j_2} 
	+ \delta_{j_1, k_2 - k_3} \delta_{k_1, j_3 - j_2} \right) 
    +  (-i)^n\left( \delta_{j_1, j_3 - k_2}\delta_{k_1 , j_2 + k_3}
	+ \delta_{j_1, k_3 - k_2} \delta_{k_1,j_2 - j_3} \right) 
\nonumber \\ [0.6em]
&& \hspace*{0.8em} \left. + \, (-1)^n
	  \left( \delta_{j_1, j_2 + j_3}\delta_{k_1, k_2 + k_3} 
	+ \delta_{j_1, j_2 - j_3}\delta_{k_1, k_2 - k_3 } 
	+ \delta_{j_1, j_2 - k_3}\delta_{k_1, j_3 + k_2}  
	+ \delta_{j_1, j_2 + k_3}\delta_{k_1, k_2 - j_3} \right)  
\rule{0em}{1em} \right]  ~, \nonumber \\
\label{bulkinteractions}
\eear
where $\overline{n}\equiv-n~{\rm mod}~4$, and we have used the fact
that $j=0$ implies $k=0$.

The quartic terms included in Eq.~(\ref{non-abelian}),
\be
-\frac{g_6^2}{4} f^{abc}f^{ade} \left( A_\mu^b A_\nu^c A^{d\,\mu} A^{e\,\nu} 
- 2 A^b_\mu A_+^c A^{d\,\mu} A_-^e - \frac{1}{2} A_+^b A_-^c A_+^d A_-^e 
\right)~,
\ee
lead to the following quartic interactions of the KK modes:
\bear
\label{quartic}
\hspace{-2.5em}
-\frac{g_4^2}{4} f^{abc} f^{ade} && \hspace{-.5em}
\left[ \rule{0mm}{5mm}
\delta_{0,0,0,0}^{(j_{1},k_{1})\cdots ( j_{4},k_{4})} 
A^{(j_{1},k_{1})\,b}_\mu A^{(j_{2},k_{2})\,c}_\nu 
A^{(j_{3},k_{3})\,d\,\mu} A^{(j_{4},k_{4})\,e\,\nu}
\right. \nonumber \\
&& + \, 2 \delta_{3,1,0,0}^{(j_{1},k_{1})\cdots ( j_{4},k_{4})} 
A^{(j_1,k_1)\,c}_+ A^{(j_2,k_2)\,e}_- 
A^{(j_3,k_3)\,b}_\mu A^{(j_4,k_4)\,d\,\mu} 
\nonumber \\ 
&& \left. 
- \, \frac{1}{2} \delta_{3,1,3,1}^{(j_{1},k_{1})\cdots ( j_{4},k_{4})} 
A^{(j_{1},k_{1})\, b}_+ A^{(j_{2},k_{2})\,c}_- A^{(j_{3},k_{3})\, d}_+ 
A^{(j_{4},k_{4})\, e}_{-}
\right]~.
\eear
Of particular interest for phenomenolgy are vertices involving at
least a zero mode, and for those one can use
Eq.~(\ref{bulkinteractions}) because
\be
\delta_{n,\overline{n},0,0}^{(j_{1},k_{1})( j_{2},k_{2} )(j_{3},k_{3})(0,0)} =
\delta_{n,\overline{n},0}^{(j_{1},k_{1})( j_{2},k_{2} )(j_{3},k_{3})} ~.
\ee

In order to extract the interactions of the physical states, the
$A_{\pm}^{(j,k)\, a}$ complex scalars have to be replaced in
Eqs.~(\ref{tri-scalar}) and (\ref{quartic}) by the two real scalar
fields, $A_H^{(j,k)\, a}$ and $A_G^{(j,k)\, a}$, as prescribed by
Eq.~(\ref{hphi}).  These heavy 4D scalars are in the adjoint
representation of the non-Abelian gauge group, so we will refer to
them as ``spinless adjoints''.

\medskip

\subsection{Ghost fields}

For completeness we now turn to determining the ghost action
associated with the gauge fixing term given by the obvious adaptation
of Eq.~(\ref{gf}) to non-Abelian fields.  This arises from inserting
in the path integral a factor
\be
\int {\cal{D}} \chi \delta[G(A)] \textrm{det}\left( 
\frac{\delta  G(A^\chi)}{\delta \chi} \right) = 1 ~,
\ee
where $\chi$ is the gauge transformation parameter, $A^{\chi}$ is the
transformed gauge field, and the gauge fixing condition is 
\be
G(A^a) = \partial^{\mu} A^{a}_{\mu} - \xi \left(\partial_4 A^{a}_4 + 
\partial_5 A^{a}_5 \right) - \omega^{a}~,
\ee
for arbitrary functions $\omega^{a}$.  After integrating with a
Gaussian weight over $\omega^{a}$ one recovers Eq.~(\ref{gf}).  Since
\be
(A^{a}_{\alpha})^{\chi} = A^{a}_{\alpha} + \frac{1}{g_6} D_{\alpha} \chi^{a}~,
\ee
with $D_{\alpha}$ the covariant derivative in the adjoint
representation, we find
\be
\frac{\delta G(A^\chi)}{\delta \chi} = \frac{1}{g_6}
\left[\partial_{\mu} D^{\mu} - 
\xi \left(\partial_{4} D_4 + \partial_{5} D_5 \right) \right] ~.
\ee
These terms in the Lagrangian may be taken into account by 
including a ghost term in the 6D action, given by
\be
-\bar{c}^a \left[ \partial_{\mu} D^{\mu} - \xi \left(\partial_{4} D_4 + 
\partial_{5} D_5 \right) \right] c^a ~,
\label{ghostaction}
\ee
where $c^a$ is an anti-commuting 6D scalar in the adjoint
representation of the gauge group.

The above procedure did not take into account the compactification of
the two extra dimensions and the associated boundary conditions.  
The free part of the 6D ghost Lagrangian (\ref{ghostaction}) is
\be
-\bar{c}^a \left[ \partial_{\mu} \partial^{\mu} - \xi \left(\partial_{4}^{2} + 
\partial_{5}^{2} \right) \right] c^a ~,
\ee
which up to the factor $\xi$ is the same as for a scalar.  It follows
that the KK expansion for the ghost fields can be written as
\be
c^a (x^{\mu}, x^{4}, x^{5}) = \frac{1}{L} \, \sum_{(j,k)}  \,
c^{(j,k)\,a}(x^{\mu}) f_{n_{c}}^{(j,k)}(x^{4}, x^{5})~,
\label{ghostexpansion}
\ee
where the $f_{n_{c}}^{(j,k)}$ belong to one of the sets of KK
functions defined in Eqs.~(\ref{KKsolns}).  The KK modes
$c^{(j,k)\,a}(x^{\mu})$ have mass $\sqrt{\xi} M_{j,k}$.  It would
appear that $n_{c}$ can take any of the integral values $0,1,2,3$.
However, for $n_{c} = 1,2,3$, the ghost fields lack the zero modes
necessary to ensure the gauge invariance of amplitudes involving the
zero-mode gauge fields.  More generally, since
$A_{\mu}(x^{\mu},x^{4},x^{5})$ satisfies boundary conditions with $n =
0$, the necessary relations between the couplings involving only gauge
fields and those involving ghosts and gauge fields can be satisfied
only if the ghosts satisfy the same boundary conditions as $A_{\mu}$.
Therefore, only the value $n_c = 0$ is allowed.  The zero mode
$c^{(0,0)}$ is the ghost required by 4D gauge invariance.

After integrating by parts Eq.~(\ref{ghostaction}), the interactions
between gauge bosons and ghost fields are given by
\be
g_6 f^{abc} (\partial^{\mu} \bar{c}^a) A_{\mu}^{b} c^{c} - 
\frac{1}{2} \xi \, g_6 f^{abc} \left[ (\partial_{+} \bar{c}^a)  A_-^{b} + 
(\partial_{-} \bar{c}^a) A_+^{b} \right] c^c~,
\label{ghostinteractions}
\ee
where $A_{\pm}$ were defined before Eq.~(\ref{Bpmbc}) and
$\partial_{\pm}$ were defined in (\ref{partialpm}).  It is worth
pointing out that the boundary conditions for $c^{a}$ given in
Eq.~(\ref{ghostexpansion}), together with the boundary conditions for
$A_{4}$, $A_{5}$ discussed in section~\ref{gaugebcs}, ensure that one
can freely integrate by parts Eq.~(\ref{ghostinteractions}) without
generating surface terms.

Using the KK expansions (\ref{ghostexpansion}), as well as
Eq.~(\ref{goldstone}), Eq.~(\ref{ghostinteractions}) leads to KK
interactions between ghosts and gauge fields:
\be
\label{ghost-coupling1}
g_4 f^{abc}  \delta_{0,0,0}^{(j_1, k_1)(j_2,k_2)(j_3,k_3)} 
\left(\partial^{\mu} \bar{c}^{(j_1,k_1)\,a}\right) A_{\mu}^{(j_2,k_2)\, b} 
c^{(j_3,k_3)\, c}~,
\ee
and between ghosts and the ``spinless adjoints'':
\be
\label{ghost-coupling2}
- \frac{i}{2} g_4 f^{abc} \xi r_{j_1,k_1} M_{j_1,k_1} 
\delta_{3,1,0}^{(j_1, k_1)(j_2,k_2)(j_3,k_3)} 
\bar{c}^{(j_1,k_1)\,a}  
A_-^{(j_2,k_2)\, b} c^{(j_3,k_3)\, c} + {\rm H.c.} ~,
\ee
where we used Eq.~(\ref{delf}) to simplify the derivative, and
$A_{\pm}^{(j,k)}$ are given in terms of the mass eigenstates
$A_{H}^{(j,k)}$ and $A_{G}^{(j,k)}$ by Eq.~(\ref{hphi}).

\bigskip

\section{Gauge interactions of fermions}
\setcounter{equation}{0}

The 6D chiral fermions have four degrees of freedom, and the Dirac
equation can be written using a set of six anti-commuting matrices,
$\Gamma^\alpha$.  We use here the following $8\times 8$
representation:
\be
\Gamma^\mu = \gamma^\mu \otimes \sigma^0 \; \; , \; \; \;
\Gamma^{4,5} = i \gamma_5 \otimes \sigma^{1,2} ~
\ee
where $\gamma^{\mu}$, are the 4D $\gamma$-matrices, $\gamma_{5}=
i\gamma^{0}\gamma^{1}\gamma^{2}\gamma^{3}$, $\sigma^0$ is the $2\times
2$ unit matrix and $\sigma^i$ are the Pauli matrices.  The 6D fermions
have $+$ or $-$ chirality, defined by the eigenvalue of the
$\overline{\Gamma} = -\gamma_5 \otimes \sigma^3$ chirality operator:
$\overline{\Gamma}\Psi_\pm = \pm\Psi_\pm$.  A 6D chiral fermion
includes both 4D chiralities: $\gamma_{5}\otimes \sigma^0 \Psi_{\pm_L}
= -\Psi_{\pm_L}$ and $\gamma_{5}\otimes \sigma^0 \Psi_{\pm_R} =
\Psi_{\pm_R}$.

It is useful to define
\be
\Gamma^{\pm} = \frac{1}{2} \left(\Gamma^{4} \pm i \Gamma^{5} \right)~,
\ee
and to observe that  
\bear
\label{Pidentities}
\Gamma^4 P_L P_\pm  = \mp i \Gamma^5 P_L P_\pm 
 ~,
\nonumber \\ [.3em] 
\Gamma^4 P_R P_\pm  = \pm i \Gamma^5 P_R P_\pm  ~,
\eear
where $P_{L,R}$, $P_{\pm}$ project on the respective 4D and 6D
chiralities, from which it follows that
\be
\Gamma^{+} \Psi_{+_L} = \Gamma^{-} \Psi_{+_R} 
= \Gamma^{-} \Psi_{-_L} = \Gamma^{+} \Psi_{-_R} =0 ~.
\ee
The fermion kinetic term can then be written in terms of 4D chiral
components as follows:
\be
\label{fermion-kinetic}
i \overline{\Psi}_{\pm} \Gamma^{\alpha} D_{\alpha} \Psi_{\pm} =
i \left(\overline{\Psi}_{\pm_L} \Gamma^{\mu} D_{\mu} \Psi_{\pm_L} + 
\overline{\Psi}_{\pm_R} \Gamma^{\mu} D_{\mu} \Psi_{\pm_R} +
\overline{\Psi}_{\pm_L} \Gamma^{\pm}   D_{\mp}   \Psi_{\pm_R} +
\overline{\Psi}_{\pm_R} \Gamma^{\mp}   D_{\pm}   \Psi_{\pm_L} \right) ~.
\ee
Here $D_{\pm} = D_4 \pm iD_5$ and $D_{\alpha}$ is the covariant
derivative defined in Eq.~(\ref{covariant}) with either $A_{\alpha}$
being an Abelian gauge field, or $A_{\alpha} = T^a A_{\alpha}^a$ where
$A_{\alpha}^a$ is a non-Abelian gauge field and $T^a$ are the gauge
group generators corresponding to the representation of $\Psi_\pm$.

The KK expansions of a 6D fermion of + chirality with
a left-handed zero mode component are \cite{Dobrescu:2004zi},
\bear
\label{KKfermionL}
\Psi_{+_L}&=& \frac{1}{L} \left[\Psi_{+_L}^{(0,0)}(x^\nu)
+ \sum_{j\ge 1} \sum_{k\ge 0} f_{0}^{(j,k)}(x^{4},x^{5}) 
\Psi_{+_L}^{(j,k)}(x^\nu) \right]
\otimes \left(\ba{c} 1 \\ 0 \ea \right)~, 
\nonumber \\ [0.5em]
\Psi_{+_R}&=& - \frac{i}{L} \sum_{j\ge 1} \sum_{k\ge 0}  
r_{j,k} \,  f_{3}^{(j,k)}(x^{4},x^{5})
\Psi_{+_R}^{(j,k)}(x^\nu) \otimes \left(\ba{c} 0 \\ 1 \ea \right)~,
\eear
while those containing a right-handed zero mode are,
\bear
\label{KKfermionR}
\Psi_{+_L}&=& \frac{i}{L} \sum_{j\ge 1} \sum_{k\ge 0}  
r_{j,k}^{*}\, f_{1}^{(j,k)}(x^{4},x^{5})
\Psi_{+_L}^{(j,k)}(x^\nu) \otimes \left(\ba{c} 1 \\ 0 \ea \right)~,
\nonumber \\ [0.5em]
\Psi_{+_R}&=& \frac{1}{L} \left[\Psi_{+_R}^{(0,0)}(x^\nu)
+ \sum_{j\ge 1} \sum_{k\ge 0} f_{0}^{(j,k)}(x^{4},x^{5}) 
\Psi_{+_R}^{(j,k)}(x^\nu) \right]
\otimes \left(\ba{c} 0 \\ 1 \ea \right) ~.
\eear
The phases of the KK functions for left- and right-handed fermions are
correlated, but one can choose an overall phase such that the
functions $f_n(x^{4},x^{5})$ with $n=0,1,3$ are given in
Eq.~(\ref{KKsolns}).  For a fermion of $-$ chirality, the same
equations are valid except for an interchange of the $L$ and $R$
indices.

After integrating the 6D fermion kinetic term shown in
Eq.~(\ref{fermion-kinetic}) over $x^4$ and $x^5$, the 4D Lagrangian
includes the usual kinetic terms for all KK modes, mass terms of the
type $M_{j,k} \overline{\Psi}^{(j,k)}_{\pm_L} \Psi^{(j,k)}_{\pm_R} +
{\rm H.c.}$, and interactions among KK modes.  The latter ones, in the
case of a fermion with $+$ chirality having a left-handed zero mode,
include interactions of the $\Psi^{(j,k)}_{+L}$ fermions with a spin-1
KK mode,
\be
\label{fermion-gauge} 
g_4 \, \delta_{0,0,0}^{(j_1, k_1)(j_2, k_2)(j_3, k_3)}\,
\overline{\Psi}^{(j_1, k_1)}_{+L} \!A^{(j_2,k_2)}_{\mu} \gamma^{\mu} 
\Psi^{(j_3,k_3)}_{+L} ~,
\ee
interactions of the $\Psi^{(j,k)}_{+R}$ fermions with a spin-1 KK mode,
\be
\label{fermion-gauge-other} 
- g_4 \, \delta_{1,0,3}^{(j_1, k_1)(j_2, k_2)(j_3, k_3)} r_{j_1,k_1}^* 
r_{j_3,k_3}
\overline{\Psi}^{(j_1,k_1)}_{+R} \!A^{(j_2,k_2)}_{\mu} \gamma^{\mu} 
\Psi^{(j_3,k_3)}_{+R} ~,
\ee
and gauge interactions of the fermions with the spinless adjoints,
\be
\label{fermion-scalar-gauge}
-i g_4 \, \delta_{0,1,3}^{(j_1,k_1)(j_2,k_2)(j_3, k_3 )}  
r_{j_2,k_2}^{*} r_{j_3, k_3} \, 
\overline{\Psi}^{(j_1,k_1)}_{+L} \left(A_H^{(j_2,k_2)} +
i A_G^{(j_2,k_2)}\right) \Psi^{(j_3,k_3)}_{+R} + {\rm H.c.}  
\ee
The $\delta_{n,\overline{n},0}^{(j_{1}k_{1})(j_2, k_2)(j_{3}k_3)}$
coefficients with $n+\overline{n} = 0$ are given in
Eq.~(\ref{bulkinteractions}), the complex numbers $r_{j,k}$ are given
in Eq.~(\ref{rjk}), and $g_{4} = g_{6}/L$ is the 4D gauge coupling.

In the case of a fermion with $+$ chirality having a right-handed zero
mode, the spin-1 KK modes have interactions with $\Psi^{(j,k)}_{+R}$
given by
\be
\label{fermion-gauge-R} 
g_4 \, \delta_{0,0,0}^{(j_1, k_1)(j_2, k_2)(j_3, k_3)}\,
\overline{\Psi}^{(j_1, k_1)}_{+R} \!A^{(j_2,k_2)}_{\mu} \gamma^{\mu} 
\Psi^{(j_3,k_3)}_{+R}  ~,
\ee
and interactions with $\Psi^{(j,k)}_{+L}$  given by
\be
\label{fermion-gauge-R-other} 
- g_4 \delta_{3,0,1}^{(j_1, k_1)(j_2, k_2)(j_3, k_3)} r_{j_1,k_1} 
r_{j_3,k_3}^*
\overline{\Psi}^{(j_1,k_1)}_{+L} \!A^{(j_2,k_2)}_{\mu} \gamma^{\mu} 
\Psi^{(j_3,k_3)}_{+L}  ~,
\ee
while the gauge interactions of the spinless adjoints with the
fermions are given by
\be
\label{fermion-scalar-gauge-R}
-i g_4 \, \delta_{0,3,1}^{(j_1,k_1)(j_2,k_2)(j_3, k_3 )}  
r_{j_2,k_2} r_{j_3, k_3}^{*} \, 
\overline{\Psi}^{(j_1,k_1)}_{+R} \left(A_H^{(j_2,k_2)} -
i A_G^{(j_2,k_2)}\right) \Psi^{(j_3,k_3)}_{+L} + {\rm H.c.}  
\ee

A 6D fermion of $-$ chirality has the same gauge interactions as
$\Psi_+$ except for an interchange of the 4D chiralities.  More
explicitly, if $\Psi_{-R}$ has a zero mode, then the gauge
interactions of $\Psi_{-R}$ and $\Psi_{-L}$ are as in
Eqs.~(\ref{fermion-gauge})-(\ref{fermion-scalar-gauge}) with an
interchange of the $L$ and $R$ indices.  If $\Psi_{-L}$ has a zero
mode, then $\Psi_{-R}$ and $\Psi_{-L}$ have gauge interactions given
by Eqs.~(\ref{fermion-gauge-R})-(\ref{fermion-scalar-gauge-R}) with
$L$ and $R$ interchanged.

\bigskip

\section{Gauge interactions of scalars}
\setcounter{equation}{0}

Consider now a 6D scalar field $\Phi$ transforming under a certain
nontrivial representation of a gauge symmetry, with an action given by
\be
S_\Phi = \int d^4 x \int_0^{\LL} dx^4 \int_0^{\LL} dx^5
\left[ \left(D_\alpha \Phi\right)^\dagger D^\alpha \Phi - M_\Phi^2 
\Phi^\dagger\Phi 
- \frac{\lambda_6}{2}\,\left(\Phi^\dagger\Phi\right)^2 \right] ~,
\label{scalar_action}
\ee
where $\lambda_6$ is a parameter of mass dimension $-2$, and
$D_{\alpha}$ is the covariant derivative associated with a gauge field
$A_{\alpha}^a$, as in Eq.~(\ref{covariant}).  As in the previous
section, we use the short-hand notation $A_{\alpha} = T^a
A_{\alpha}^a$ where $T^a$ are the gauge group generators corresponding
to the representation of $\Phi$.

The KK decomposition of the scalar has been derived in
\cite{Dobrescu:2004zi}:
\be
\Phi (x^\mu,x^4,x^5) = \frac{1}{L} \sum_{(j,k)} \Phi^{(j,k)}(x^\mu) 
f_n^{(j,k)}(x^{4},x^{5})~,
\ee
with $n = 0,1,2$, or $3$. The scalar KK modes $\Phi^{(j,k)}$ have masses
\be
M^{(j,k)}_\Phi = \sqrt{M_\Phi^2 + M_{j,k}^2} ~,
\label{fimass}
\ee
where $M_{j,k}$ are the KK masses given in Eq.~(\ref{massdef}).

Using the KK decomposition of gauge fields given in
Eq.~(\ref{KKgaugepm}), and integrating over the extra dimensional
coordinates, we find that the 4D Lagrangian includes interactions of
two KK scalars with one spin-1 KK field,
\be
\label{amuphi}
i g_4 \delta_{\overline{n},0,n}^{(j_{1},k_{1})(j_{2},k_{2})(j_{3},k_{3})}
\Phi^{(j_{1},k_{1})\dag} A_{\mu}^{(j_{2},k_{2})} 
\partial^{\mu} \Phi^{(j_{3},k_{3})} + {\rm H.c.} ~,
\ee
as well as interactions of two KK scalars with two spin-1 KK fields,
\be
\label{amuanuphi}
g_4^2 \delta_{\overline{n},0,0,n}^{(j_{1},k_{1})...(j_{4},k_{4})} 
\Phi^{(j_{1},k_{1})\dag} A_{\mu}^{(j_{2},k_{2})} A^{(j_{3},k_{3})\,\mu} 
\,\Phi^{(j_{4},k_{4})} ~.
\ee
In particular, the interactions of the $A_{\mu}^{(0,0)}$ fields are 
dictated by 4D gauge invariance.

The 4D Lagrangian also includes interactions of two $\Phi^{(j,k)}$
scalars with one of the $A_H^{(j,k)}$ and $A_G^{(j,k)}$ spinless
adjoints,
\bear
\label{ahoragphi}
&& \hspace{-7mm}
\frac{g_4}{2} M_{j_3,k_3} 
\Phi^{(j_{1},k_{1})\dag}
\left[ \rule{0pt}{17pt}
\left(
\delta_{\overline{n},1,n-1}^{(j_{1},k_{1})(j_{2},k_{2})(j_{3},k_{3})}\,
r_{j_2,k_2}^{*}\,r_{j_3,k_3}
- \delta_{\overline{n},3,n+1}^{(j_{1},k_{1})(j_{2},k_{2})(j_{3},k_{3})}\, 
r_{j_2,k_2}\,r_{j_3,k_3}^* 
\right)
 A_H^{(j_2,k_2)}  \right. 
\\ [1em]
&& \left. \hspace{-5mm}
\mbox{} + i \left(\delta_{\overline{n},3,n+1}^{(j_{1},k_{1})(j_{2},k_{2})(j_{3},k_{3})}\,
r_{j_2,k_2}\,r_{j_3,k_3}^* +
\delta_{\overline{n},1,n-1}^{(j_{1},k_{1})(j_{2},k_{2})(j_{3},k_{3})}\,
r_{j_2,k_2}^{*}\,r_{j_3,k_3}\right)  A_G^{(j_{2},k_{2})} 
\right] \Phi^{(j_3,k_3)} + {\rm H.c.} ~,
\nonumber
\eear
[here we replaced the derivatives using Eq.  (2.17)], interactions of
two $\Phi^{(j,k)}$ scalars with two of the spinless adjoints,
\be
g_4^2 r_{j_2,k_2}r_{j_3,k_3}^*
\delta_{\overline{n},3,1,n}^{(j_{1},k_{1})\cdots (j_4,k_4)}\,
\Phi^{(j_{1},k_{1})\dag}\left( A_H^{(j_2,k_2)} - iA_G^{(j_2,k_2)} \right)
\left( A_H^{(j_3,k_3)} + iA_G^{(j_3,k_3)} \right) \Phi^{(j_4,k_4)} ~,
\label{ahagphi}
\ee
and, finally, self-interactions of the $\Phi^{(j,k)}$ scalars
\be
- \frac{\lambda_4}{2}\delta_{\overline{n},n,\overline{n},n}^{(j_{1},k_{1})
\ldots (j_4,k_4)}\,
\Phi^{(j_{1},k_{1})\dag}\Phi^{(j_{2},k_{2})} \Phi^{(j_{3},k_{3})\dag} 
\Phi^{(j_4,k_4)} ~,
\label{self}
\ee
where $\lambda_4=\lambda_6/L^2$ is the 4D quartic coupling.

\section{Spontaneous symmetry breaking}
\setcounter{equation}{0}

We now discuss the case where the gauge symmetry is broken by the VEV
of a 6D scalar $\Phi$.  The action is given in
Eq.~(\ref{scalar_action}), with $M_\Phi^2 <0$.  By adding an
irrelevant constant, the terms in the 6D Lagrangian that include
$\Phi$ can be rewritten as
\be
{\cal L}_\Phi = \left|D_{\alpha}\Phi\right|^{2} 
- \frac{\lambda_6}{2}\,
\left( \Phi^\dagger \Phi - \frac{1}{2} \vs^{2} \right)^2 ~,
\label{SSBPot}
\ee
where $\vs > 0$ is the 6D VEV, which has mass dimension two and is
defined by the relation
\be
- M_{\Phi}^{2} = \frac{1}{2} \lambda_{6} \vs^{2}~.
\ee

Let us focus on the case where the gauge group is $SU(N)$ and the
scalar $\Phi$ is in the fundamental representation of the gauge group.
The formulae we derive below also apply (with trivial modifications)
to $SO(N)$ or $U(1)$ gauge groups.  In order to analyze the spectrum
and interactions in the presence of the 6D VEV, we parameterize $\Phi$
as
\be
\Phi = \frac{1}{\sqrt{2}} (\vs + h ) \, U_\eta \vev~,
\ee
where $\vev = (0,\ldots,0,1)$ defines the direction of the VEV, $h$ is
the single 6D real scalar that is orthogonal to the Nambu-Goldstone
modes, and $U_\eta$ is an unitary matrix that depends on the 6D
Nambu-Goldstone bosons, $\eta^a$: \be U_{\eta} = e^{i \eta/\vs} ~, \ee
where $\eta = \eta^{a} X^{a}$, with $X^a$ the broken generators.  The
sum over $a$ includes the $2N-1$ generators of the coset
$SU(N)/SU(N-1)$.
In terms of $h$ and $\eta$, the Lagrangian (\ref{SSBPot}) is given by
\be
{\cal L}_\Phi = \frac{1}{2} ( \partial_{\alpha} h )^{2} 
- \frac{\lambda_6}{8}\, h^2
\left( 2 \vs + h \right)^2
+ \frac{1}{2} ( \vs + h )^{2} \, \vev^{\dagger} \left|
U_{\eta}^{\dagger} \partial_{\alpha} U_{\eta} 
- i g_{6} U_{\eta}^{\dagger} A_{\alpha} U_{\eta} \right|^{2} \vev ~.
\label{cov}
\ee
where $A_\alpha = A_\alpha^a T^a$, with $a$ running over all group
generators $T^a$.

\medskip

\subsection{Physical states}

Expanding $U_\eta$ in powers of $\eta$, we find the terms in ${\cal
L}_\Phi$ that are quadratic in $\eta$ or $A_\alpha$:
\be
\frac{1}{2} \vs^{2} \, \vev^{\dagger} \left( \frac{1}{\vs} 
\partial_{\alpha} \eta - g_{6} A_{\alpha} \right)^{2} \vev ~.
\label{cov-exp}
\ee
After integration by parts, these become
\be
\frac{1}{2} \vev^{\dagger} \left[
(\partial_{\alpha} \eta)^{2} + g_{6} \vs \left\{ \eta , \partial_{\alpha} 
A^{\alpha} \right\} + g_{6}^{2} \vs^{2} 
A_{\alpha} A^{\alpha}
\right] \vev  ~,
\label{quadraticSSB}
\ee
where $\{...  , ...  \}$ is the anticommutator.  We see that
$\eta^{a}$ mixes with the broken components of both $A^{a}_{\mu}$ and
$\partial_{4} A^{a}_{4} + \partial_{5} A^{a}_{5}$, where the latter
includes the KK modes $ A_{G}^{(j,k) \, a}$ that are eaten by the
spin-1 modes in the limit of zero VEV [see Eq.~(\ref{goldstone})].  In
order to make the physics more transparent, we modify the gauge fixing
term (\ref{gf}) to
\be
{\cal L}^{v}_{GF}=-\frac{1}{2\xi}
\left[\frac{}{} \partial_\mu A^{\mu \, a} - \xi\left(\partial_4A^{a}_4 + 
\partial_5 A^{a}_5 \right) + \xi g_{6} \vs \eta^{a} \right]^2~,
\label{gfv}
\ee
where it is understood that $\eta^{a}$ is non-vanishing only along the
direction of the broken generators.  The terms involving the unbroken
components of $A_{\mu}^{a}$ remain as in Eq.~(\ref{gf}).  Since
$\vev^{\dagger} \left\{ \eta, \partial_{\alpha} A^{\alpha} \right\}
\vev = 2 \eta^{a} \partial_{\alpha} A^{\alpha \, a}$, where $a$ runs
over the broken generators, the crossed terms in Eq.~(\ref{gfv})
involving $\eta^{a}$ and $A^{a}_{\mu}$ cancel the corresponding terms
in (\ref{quadraticSSB}).  The remaining terms in
Eqs.~(\ref{quadraticSSB}) and (\ref{gfv}), involving $\eta$, $A_{4}$
and $A_{5}$, are then
\be
\frac{1}{2} \vev^{\dagger} \left[ (\partial_{\alpha} \eta)^{2} - 
g_{6}^{2} \vs^{2} A_{+} A_{-} 
- \xi \, g_{6}^{2} \vs^{2} \eta^{2} + g_{6} \vs (\xi - 1) 
\{ \eta, \partial_{4} A_{4} + \partial_{5} A_{5} \} \right] \vev~,
\label{etaAG_action}
\ee
together with the second line in Eq.~(\ref{s2}).  Here we used
$A^{a}_{\alpha} A^{\alpha \, a} = A^{a}_{\mu} A^{\mu \, a} - A^{a}_{+}
A^{a}_{-}$, where $A^{a}_{\pm} = A_4 \pm i A_5$.

We turn next to the KK decomposition.  This is achieved by using the
KK expansions for $A^{a}_{\mu}$ and $A^{a}_{\pm}$,
Eqs.~(\ref{KKgaugepm}), supplemented by
\bear
h(x^\mu,x^4,x^5) &=& \frac{1}{L} \sum_{(j,k)} h^{(j,k)}(x^\mu) 
f_0^{(j,k)}(x^{4},x^{5})~,
\nonumber \\
\eta^{a}(x^\mu,x^4,x^5) &=& \frac{1}{L} \sum_{(j,k)} \eta^{(j,k) \, a}(x^\mu) 
f_0^{(j,k)}(x^{4},x^{5})~.
\label{hetaexpansions}
\eear
In assuming a constant VEV for $\Phi$ we have implicitly imposed that
it satisfy boundary conditions with $n=0$, a fact we used in
Eqs.~(\ref{hetaexpansions}).

Using Eqs.~(\ref{goldstone}) as well as the orthogonality of the KK
functions $f_n^{(j,k)}$, it is now straightforward to obtain the new
terms in the KK action.  One finds new mass contributions proportional
to the VEV for the broken components of $A^{(j,k)\, a}_\mu$ and
$A^{(j,k)\,a}_H$ modes, where the latter ones are defined in
Eq.~(\ref{hphi}):
\be
- \frac{1}{2} g_{4}^{2} \vf^{2} \, \vev^{\dagger} 
\left( A^{(j,k)}_{\mu} A^{(j,k)\,\mu} + 
A^{(j,k)}_{H} A^{(j,k)}_{H} \right) \vev~.
\ee
These, together with the mass contributions due to momentum in extra
dimensions, shown in Eqs.~(\ref{spin_one_action}) and (\ref{l4dfin}),
lead to masses for these modes given by
\be
M_A^{(j,k)} = \sqrt{ M_{j,k}^{2} + g_{4}^{2} \vf^{2} }~,
\label{newMjk}
\ee
with $M_{j,k}$ defined by Eq.~(\ref{massdef}).  We chose to express
these results in terms of the 4D gauge coupling, $g_{4} = g_{6}/L$,
and the 4D VEV, $\vf = \vs L$, with mass dimension one.  Also, from
Eq.~(\ref{cov}), the masses of the $h^{(j,k)}$ real scalars are given
by
%
\be
M_h^{(j,k)} = \sqrt{M^2_{j,k} + \lambda_{4} \vf^{2}}~,
\ee
where $\lambda_{4} = \lambda_{6}/L^{2}$ is the 4D quartic coupling.

Finally, the mass terms involving $\eta^{(j,k)}$ and $A_{G}^{(j,k)}$
[see Eqs.~(\ref{etaAG_action}) and (\ref{l4dfin})] mix these modes,
such that, for the components along the broken generators, the
physical states are given by the linear combinations
\bear
\tilde{A}_{G}^{(j,k) \, a} &=& \frac{1}{M_A^{(j,k)}} 
\left( M_{j,k} A_{G}^{(j,k) \, a}
+ g_{4} \vf \, \eta^{(j,k) \, a} \right)~,
\nonumber \\
\tilde{\eta}^{(j,k) \, a} &=& \frac{1}{M_A^{(j,k)}} 
\left( M_{j,k} \eta^{(j,k) \, a}
- g_{4} \vf A_{G}^{(j,k) \, a} \right)~,
\label{AGchinew}
\eear
for $j > 0$ and $k \geq 0$, while for $j=k=0$ we define
\be
\tilde{A}_{G}^{(0,0) \, a} = \eta^{(0,0) \, a}~.
\label{zero-NGB}
\ee
This latter mode is eaten by the would-be zero-mode of the gauge KK
tower.  The free Lagrangian terms for these states are
\be
\frac{1}{2} \vev^{\dagger} \left[ \left(\del_\mu \tilde{\eta}^{(j,k)}\right)^2 
- \left( M_A^{(j,k)} \tilde{\eta}^{(j,k)}\right)^2
+\left(\del_\mu \tilde{A}_G^{(j,k)}\right)^2 -\xi
M_{(j,k)}^{2} \left(\tilde{A}_G^{(j,k)}\right)^2\right] \vev ~,
\label{AGchi_action}
\ee
while for the unbroken components it is given by
\be
\frac{1}{2} \left[ \left(\del_\mu A_G^{(j,k) \, a}\right)^2 -\xi
M_{j,k}^{2} \left( A_G^{(j,k) \, a}\right)^2\right] ~.
\ee

Note that under an infinitesimal 6D gauge transformation 
%
\be
A^{a}_{\alpha} \mapsto A^{a}_{\alpha} + \frac{1}{g_{6}} 
\partial_{\alpha} \chi^{a} + \cdots ~,
\hspace{1cm} \eta^{a} \mapsto \eta^{a} + \vs \chi^{a} + \cdots~,
\ee 
one has $\partial_{4} A^{a}_{4} + \partial_{5} A^{a}_{5} \mapsto
\partial_{4} A^{a}_{4} + \partial_{5} A^{a}_{5} + (\partial_{4}^{2} +
\partial_{5}^{2}) \chi^{a} /g_{6} + \cdots$, which after KK expansion,
Eq.~(\ref{goldstone}), translates at the linear level into
\be
A_{G}^{(j,k)} \mapsto A_{G}^{(j,k)} + \frac{1}{g_{4}} M_{j,k} \chi^{(j,k)} + \cdots~,
\hspace{1cm} \eta^{(j,k)} \mapsto \eta^{(j,k)} + \vf \chi^{(j,k)} + \cdots~.
\ee
In the above, the dimensionless gauge transformation parameter, $\chi$, has
an expansion
\be
\chi(x^\mu,x^4,x^5) = \sum_{(j,k)} \chi^{(j,k)}(x^\mu) 
f_0^{(j,k)}(x^{4},x^{5})~,
\label{chiexpansion}
\ee
and the $\ldots$ stand for higher order terms that, in general, mix
the KK levels.  It is then clear that the inhomogeneus piece of the
gauge transformation drops from $\tilde{\eta}^{(j,k) \, a}$, while the
orthogonal combination, $\tilde{A}_{G}^{(j,k) \, a}$, shifts under the
gauge transformation.  $\tilde{A}_{G}^{(j,k)}$ is the generalization
of the would-be Nambu-Goldstone boson of sections 2 and 3 to the case
where the gauge symmetry is broken by a constant scalar VEV. Note that
in the unitary gauge, $\xi \rightarrow \infty$, each KK level gauge
boson eats a linear combination of $A_{4}$, $A_{5}$ and $\eta$.

As a result of the modification arising from $\vs$ in
the gauge fixing term, Eq.~(\ref{gfv}), the ghost Lagrangian,
Eq.~(\ref{ghostaction}), receives a new contribution
\be
- \xi g_{6}^{2} \vs^{2} \, \bar{c}^a c^a ~,
\ee
where $a$ runs over the broken components only.  Hence, in the
presence of the Higgs VEV, the KK masses associated with the broken
components of the ghost fields are $\sqrt{\xi} M_A^{(j,k)}$, where
$M_A^{(j,k)}$ was defined in Eq.~(\ref{newMjk}), while those
associated with the unbroken ones remain at $\sqrt{\xi} M_{j,k}$.

\medskip

\subsection{Trilinear and quartic couplings}

The interactions among KK modes follow from the nonlinear terms in
Eq.~(\ref{cov}).  These can be derived by expanding $U_\eta$ in a
power series:
\bear
U_{\eta}^{\dagger} \partial_{\alpha} U_{\eta} &=& i \partial_{\alpha} \etahat
- \frac{1}{2} \left[ \partial_{\alpha} \etahat, \etahat \right]
- \frac{i}{3!} \left[ \left[ \partial_{\alpha} \etahat, \etahat \right] , 
\etahat \right] + \cdots ~,
\nonumber \\
U_{\eta}^{\dagger} A_{\alpha} U_{\eta} &=& 
A_{\alpha} + i \left[ A_{\alpha} , \etahat \right]
- \frac{1}{2} \left[ \left[ A_{\alpha}, \etahat \right], \etahat \right]
- \frac{i}{3!} \left[ \left[ \left[ A_{\alpha}, \etahat \right], \etahat \right], \etahat \right]  + \cdots ~,
\label{matrixidentities}
\eear
where $[...,...]$ is the commutator and $\etahat = \eta/\vs$.

The interaction terms up to quadratic order in $\eta$ and $A_\alpha$
that appear in the Lagrangian ${\cal L}_\Phi$ are 
\be
- \frac{\lambda_6}{8} \left( 4 \vs h^{3} + h^{4} \right) +
\frac{1}{2} \left(2 \vs h + h^{2} \right) \, \vev^{\dagger} 
\left( \partial_{\alpha} \etahat - g_{6} A_{\alpha} 
\right)^{2} \vev ~.
\label{interactions1}
\ee
Higher order terms in $\eta$ and $A_\alpha$ include additional
trilinear and quartic interactions involving the $h$ scalar,
\be
\frac{i}{2} \vs \left(\vs +2 h\right) \, \vev^{\dagger} 
\left\{ \frac{1}{2} 
\left\{ \partial_{\alpha} \etahat, \left[ \partial^{\alpha} 
\etahat, \etahat \right] \right\}
+ g_{6} \left[ \etahat \, (\partial_{\alpha} \etahat ) A^{\alpha}
- (\partial_{\alpha} \etahat ) A^{\alpha} \etahat  \right]
\right\} \vev~,
\label{higherorder1}
\ee
interactions of three $\eta$ fields with an $A_\alpha$,
\be
\frac{g_{6}}{2} \vs^2 \, \vev^{\dagger} 
\left[ \rule{0pt}{17pt}
(\partial_{\alpha} \etahat ) \left[ A^{\alpha} , \etahat \right] \etahat 
- \etahat \left[ A^{\alpha} , \etahat \right] \partial_{\alpha} \etahat 
+ \etahat (\partial_{\alpha} \etahat ) \left[ \etahat, A^{\alpha} \right]
- \left[ \etahat, A^{\alpha} \right] (\partial_{\alpha} \etahat )\, \etahat
\right] \vev~,
\label{higherorder2}
\ee
quartic interactions among the $\eta$ fields,
\be
- \frac{1}{2}\vs^2 \, \vev^{\dagger} 
\left\{ \rule{0pt}{17pt} \frac{1}{3!} \left(\rule{0pt}{13pt}
(\partial_{\alpha} \etahat )
\left[ \left[ \partial_{\alpha} \etahat, \etahat \right] , 
\etahat \right] + {\rm H.c.} \right)
+ \frac{1}{4} \left[ \partial_{\alpha} \etahat, \etahat \right]^2
\right\}\vev~,
\label{higherorder3}
\ee
and interactions involving five or more fields.

Replacing the KK expansions, Eq.~(\ref{interactions1}) leads to
trilinear interactions involving the spin-1 modes
\be
\label{SSB_trilinear_1}
\vf \, \delta_{0,0,0}^{(j_{1},k_{1})(j_{2},k_{2})(j_{3},k_{3})}
h^{(j_{1},k_{1})} \vev^{\dagger} \left( \frac{1}{\vf} \partial_{\mu} 
\eta^{(j_{2},k_{2})} - g_{4} A_{\mu}^{(j_{2},k_{2})} \right) 
\left( \frac{1}{\vf} \partial^{\mu} \eta^{(j_{3},k_{3})} - 
g_{4} A^{(j_{3},k_{3}) \, \mu} \right) \vev
~.
\ee
Additional trilinear couplings of a spin-1 mode, coming from
Eq.~(\ref{higherorder1}), are given by
\be
\frac{i}{2}g_4  \, \delta_{0,0,0}^{(j_{1},k_{1})(j_{2},k_{2})(j_{3},k_{3})}
\, \vev^{\dagger} 
\left[ \eta^{(j_{1},k_{1})} \, \left(\partial^{\mu} \eta^{(j_{2},k_{2})} 
\right) 
A^{(j_{3},k_{3})}_\mu
- \left(\partial^{\mu} \eta^{(j_{1},k_{1})} \right) A^{(j_{2},k_{2})}_\mu 
\eta^{(j_{3},k_{3})}  \right]
 \vev~,
\ee
In the above two equations, one should express $\eta^{(j,k)}$ in terms
of the mass eigenstates $\tilde{\eta}^{(j,k)}$ and
$\tilde{A}_{G}^{(j,k)}$ by inverting Eqs.~(\ref{AGchinew}):
\be
\eta^{(j,k) \, a} = \frac{1}{M_A^{(j,k)}} \left( M_{j,k} 
\tilde{\eta}^{(j,k) \, a} + g_{4} \vf \tilde{A}_{G}^{(j,k) \, a} \right)~,
\ee
except for the zero-mode, which is given by Eq.~(\ref{zero-NGB}).
There are also trilinear interactions among scalars which include at
least one $h^{(j,k)}$ field and no derivatives:
\bear
\label{SSB_trilinear_2}
\vf \, h^{(j_{1},k_{1})} 
&& \hspace*{-0.5em}  \left\{ \rule{0mm}{5mm} 
 - \frac{1}{2} \lambda_{4} \, 
\delta_{0,0,0}^{(j_{1},k_{1})(j_{2},k_{2})(j_{3},k_{3})} 
h^{(j_{2},k_{2})} h^{(j_{3},k_{3})} 
+ r_{j_2,k_2} r_{j_3,k_3}^{*}  
\delta_{0,3,1}^{(j_{1},k_{1})(j_{2},k_{2})(j_{3},k_{3})} 
\right. \nonumber \\ [0.3em]
&& \left. \;\; \times \,
\vev^{\dagger} \left( g_{4} A_{H}^{(j_{2},k_{2})} +  i \omega_{j_{2},k_{2}} 
\tilde{\eta}^{(j_{2},k_{2})} \right) 
\left( g_{4} A_{H}^{(j_{3},k_{3})} -  i \omega_{j_{3},k_{3}} 
\tilde{\eta}^{(j_{3},k_{3})} \right) \vev
\rule{0mm}{5mm} \right\}~,
\eear
where we defined the KK-number-dependent, dimensionless ratios
\be
\omega_{j,k} = \frac{M_A^{(j,k)}}{\vf}~.
\ee
The remaining trilinear couplings involve only $\tilde{\eta}^{(j,k)}$,
$A_{H}^{(j,k)}$ and $A_G^{(j,k)}$ scalars, and can be derived from
Eq.~(\ref{higherorder1}).

The quartic interactions include terms involving $h^{(j,k)}$ and the
spin-1 fields,
\be
\label{SSB_quartic_1}
\frac{1}{2} \delta_{0,0,0,0}^{(j_{1},k_{1})\cdots(j_{4},k_{4})}
h^{(j_{1},k_{1})} h^{(j_{2},k_{2})} \vev^{\dagger} \left( \frac{1}{\vf} 
\partial_{\mu} \eta^{(j_{3},k_{3})} - g_{4} A_{\mu}^{(j_{3},k_{3})} \right) 
\left( \frac{1}{\vf} \partial^{\mu} \eta^{(j_{4},k_{4})} - g_{4} 
A^{(j_{4},k_{4}) \, \mu} \right) \vev
~,
\ee
as well as the spinless fields $h^{(j,k)}$, $A_{H}^{(j,k)}$ and
$\tilde{\eta}^{(j,k)}$:
\bear
\label{SSB_quartic_2}
h^{(j_{1},k_{1})} h^{(j_{2},k_{2})} 
&& \hspace*{-0.5em} \left\{ \rule{0mm}{5mm} 
- \frac{1}{8} \lambda_{4} \, 
\delta_{0,0,0,0}^{(j_{1},k_{1}) \cdots (j_{4},k_{4})} 
h^{(j_{3},k_{3})} h^{(j_{4},k_{4})} 
+ r_{j_3,k_3} r_{j_4,k_4}^{*}  
\delta_{0,0,3,1}^{(j_{1},k_{1}) \cdots (j_{4},k_{4})} 
\mbox{} \hspace{2.6cm}
\right. \nonumber \\ [0.3em]
&& \left. \; \times \, 
\vev^{\dagger} \left( g_{4} A_{H}^{(j_{3},k_{3})} +  i \omega_{j_{3},k_{3}} 
\tilde{\eta}^{(j_{3},k_{3})} \right) 
\left( g_{4} A_{H}^{(j_{4},k_{4})} -  i \omega_{j_{4},k_{4}} 
\tilde{\eta}^{(j_{4},k_{4})} \right) \vev
\rule{0mm}{5mm} \right\}~.
\eear
The higher order terms in Eqs.~(\ref{higherorder1}),
(\ref{higherorder2}) and (\ref{higherorder3}) lead to additional
quartic couplings among KK modes, which include at most a single 4D
spin-1 field.  These couplings are straightforward to derive but have
long expressions so that we do not display all of them here.

We end this section by observing that the Abelian case can be
recovered from the previous formulae by setting $\vev = 1$, and
considering a single gauge index $a$.

\bigskip

\section{Conclusions} \setcounter{equation}{0}

Gauge theories in six dimensions may be relevant for physics beyond
the Standard Model provided the size of two dimensions is below
$10^{-16}$ {\rm cm}.  Theories of this type have been proposed in the
past, with compactification scales ranging from the electroweak scale
to the GUT scale.  This paper presents the first in-depth study of the
gauge interactions among KK modes.  We have concentrated on the
simplest compactification of two dimensions that leads to zero mode
fermions of definite 4D chirality, namely the ``chiral square''
defined in Ref.~\cite{Dobrescu:2004zi}.

After identifying a gauge fixing procedure appropriate for this
compactification, we have determined a set of gauge invariant boundary
conditions for gauge fields.  The ensuing KK decomposition of a 6D
gauge field $A^\alpha$, $\alpha = 0,1, \ldots, 5$, includes a tower of
spin-1 modes, $A_\mu^{(j, k)}$, that have a zero mode ($j=k=0$), and
two towers of spin-0 modes that have no zero mode, $A_4^{(j, k)}$ and
$A_5^{(j, k)}$, where the pair of KK numbers $(j,k)$ take the integer
values $j \ge 1$, $k \ge 0$.  The spin-1 zero mode is associated with
the unbroken 4D gauge invariance, while the other spin-1 modes become
heavy.  Each nonzero spin-1 mode, $A_\mu^{(j, k)}$, has a longitudinal
polarization given by a linear combination of $A_4^{(j, k)}$ and
$A_5^{(j, k)}$.  The linear combination $A_G^{(j, k)}$ of spin-0 modes
that play the role of Nambu-Goldstone boson eaten by the heavy spin-1
mode depends on the KK numbers, as shown in Eq.~(\ref{hphi}).  The
orthogonal combination, $A_H^{(j, k)}$, is gauge invariant and remains
as an additional physical degree of freedom.  Therefore, unlike the
case of one extra dimension, where the extra components of the gauge
field can be gauged away, any gauge field in two extra dimensions
implies the existence of a tower of heavy spinless particles in the
adjoint representation of the gauge group (``spinless adjoints''),
whose interactions depend on the KK numbers.

The self-interactions of 6D non-Abelian gauge fields induce the
following terms in the 4D Lagrangian involving KK modes: trilinear
couplings of spin-1 modes, given in Eq.~(\ref{tri-gauge}); couplings
of two spin-1 modes and one spin-0 mode, and couplings of three spin-0
modes, given in Eq.~(\ref{tri-scalar}); quartic couplings of spin-1
modes, couplings of two spin-1 modes and two spin-0 modes, and quartic
couplings of spin-0 modes, given in Eq.~(\ref{quartic}); and finally,
couplings of one spin-1 mode, or one spinless adjoints, and two modes 
of the ghost field, given in Eqs.~(\ref{ghost-coupling1}) and (\ref{ghost-coupling2}), respectively.

The gauge interactions of a chiral 6D fermion induce couplings of a
gauge field mode to fermion modes of both 4D chiralities.  These
depend on the 4D chirality of the zero-mode fermion.  The spin-1 modes
couple to fermion modes according to Eqs.~(\ref{fermion-gauge}),
(\ref{fermion-gauge-other}), (\ref{fermion-gauge-R}), and
(\ref{fermion-gauge-R-other}), while the Yukawa couplings of the
spinless adjoints to the fermion modes are given in
Eqs.~(\ref{fermion-scalar-gauge}) and (\ref{fermion-scalar-gauge-R}).

The gauge interactions of a 6D scalar field induce couplings of two
scalar modes to one or two spin-1 modes, given in Eq.~(\ref{amuphi})
and (\ref{amuanuphi}), and also to one or two spinless adjoints, given
in Eqs.~(\ref{ahoragphi}) and (\ref{ahagphi}).  A 6D quartic
self-interaction of a scalar induces quartic couplings of the scalar
modes, as in Eq.~(\ref{self}).

We have also studied the case where the gauge symmetry is broken by a
the VEV of a 6D scalar that has a zero mode. In this case, all gauge KK 
modes receive a
contribution to their mass from the spontaneous breaking.  The
longitudinal polarizations of the heavy KK modes are now given by a
linear combination of $A_{4}$, $A_{5}$ and the scalar that acquires
the VEV, as shown in Eq.~(\ref{AGchinew}).  The longitudinal
polarization of the would-be zero-mode gauge field
is provided by the zero-mode of the additional spinless adjoint, as
given in Eq.~(\ref{zero-NGB}).  We showed how the longitudinal modes
and additional scalars can be identified by studying the
transformation properties under 6D gauge transformations.  We
displayed the interactions of these scalars with the spin-1 modes and
among themselves in
Eqs.~(\ref{SSB_trilinear_1})-(\ref{SSB_quartic_2}).

All the couplings among various KK modes mentioned above are induced
at tree level by bulk interactions.  Loop corrections generate
operators localized at the three conical singularities of the chiral
square.  Even though these preserve KK parity, they lead to additional
couplings among KK modes.  These are perturbative corrections, but
nevertheless may have important phenomenological consequences as they
give rise to mixing among all KK modes belonging to the same tower.
This effect is studied in a subsequent paper \cite{us1}.

The detailed construction of 6D gauge theories with explicit boundary
conditions presented here opens up various theoretical and model
building avenues of research, including issues pertaining to symmetry
breaking by boundary conditions \cite{Hebecker:2001jb}, relation to
other compactifications \cite{Scrucca:2003ut}, the structure of gauge
and gravitational anomalies on a compact space
\cite{Gherghetta:2002nq}, and latticized or deconstructed
\cite{Hill:2000mu} versions of the chiral square compactification.

Given the constraints on the compactification scale of universal extra
dimensions from electroweak measurements \cite{Appelquist:2000nn},
flavor-changing processes \cite{Agashe:2001xt} or collider searches
\cite{Cheng:2002iz, Rizzo:2001sd},
are as low as a few hundred GeV, it would be particularly interesting
to use the tools developed here for analyzing the phenomenological
implications of the Standard Model in two universal extra dimensions
compactified on the chiral square.  That model is well motivated by
the natural happenstances of proton stability, constraint on the
number of fermion generations, and existence of a dark matter
candidate.  More generally, the spectra and interactions derived here
are relevant for any extensions of the Standard Model in the context
of 6D theories.

\bigskip

{\bf Acknowledgements:} \ We have benefited from the hospitality
of the Aspen Center for Physics at various stages of this project.
G.B. acknowledges the support of the State of S\~{a}o Paulo's Agency for the 
Promotion of Research (FAPESP), and the Brazilian National Council for
Technological and Scientific Development (CNPq).
The work of B.D. was supported by DOE under
contract DE-FG02-92ER-40704. 


 \vfil \end{document}